\documentclass[a4paper,fleqn]{cas-sc}

\usepackage[numbers]{natbib}
\usepackage{caption}
\usepackage{makecell}
\usepackage{graphicx}
\usepackage{float}
\usepackage{subfigure}
\usepackage{multirow}
\usepackage{varioref}
\labelformat{figure}{Fig~#1}
\labelformat{table}{Table~#1}
\usepackage{hyperref}
\usepackage{subfigure}
\usepackage{bm}
\usepackage{amsmath}
\usepackage[ruled,vlined,linesnumbered]{algorithm2e} 

\usepackage{soul}
\usepackage{amsthm} 

\usepackage{caption}
\usepackage{makecell}
\usepackage{graphicx}
\usepackage{float}
\usepackage{subfigure}
\usepackage{multirow}
\usepackage{varioref}
\usepackage{subfigure}
\usepackage{bm}
\usepackage{amsmath}
\usepackage[ruled,vlined,linesnumbered]{algorithm2e} 
\usepackage{soul}
\usepackage{booktabs} 
\usepackage{array}

\newtheorem{myDef}{Definition}

\hypersetup{colorlinks = blue,
	linkcolor=blue,
	filecolor = blue,
	urlcolor = blue,
	anchorcolor=blue,
	citecolor=blue}

\usepackage{threeparttable}

\def\tsc#1{\csdef{#1}{\textsc{\lowercase{#1}}\xspace}}
\tsc{WGM}
\tsc{QE}
\tsc{EP}
\tsc{PMS}
\tsc{BEC}
\tsc{DE}

\begin{document}
\captionsetup[figure]{labelfont={bf},labelformat={default},labelsep=period,name={Fig.}}
\let\WriteBookmarks\relax
\def\floatpagepagefraction{1}
\def\textpagefraction{.001}
\let\printorcid\relax

\shorttitle{}

\shortauthors{}

\title [mode = title]{Toward Trustworthy Identity Tracing via Multi-attribute Synergistic Identification}

\author[3]{Decheng Liu}
[orcid=0000-0002-6550-212X]
\cormark[1]

\author[1,2]{Jiahao Yu}
\cormark[1]
\cortext[cor1]{Equal contributions.}

\author[1,3]{Ruimin Hu}
[orcid=0000-0002-5872-3872]
\cormark[2]
\cortext[cor2]{Corresponding author: hrm1964@163.com}

\author[1,2]{Wenbin Feng}

\credit{Data curation, Writing - Original draft preparation}

\address[1]{National Engineering Research Center for Multimedia Software, School of Computer Science, Wuhan University, Wuhan 430072, China}
\address[2]{Hubei Key Laboratory of Multimedia and Network Communication Engineering, Wuhan University, Wuhan 430072, China}
\address[3]{School of Cyber Engineering, Xidian University, Xi’an 710071, China}

\begin{abstract}
Identity tracing is a technology that uses the selection and collection of identity attributes of the object to be tested to discover its true identity, and it is one of the most important foundational issues in the field of social security prevention.
However, traditional identity recognition technologies based on single attributes have difficulty achieving ultimate recognition accuracy, where deep learning-based model always lacks interpretability. 
Multivariate attribute collaborative identification is a possible key way to overcome mentioned recognition errors and low data quality problems.
In this paper, we propose the “Trustworthy Identity Tracing” (TIT) task and a Multi-attribute Synergistic Identification based TIT framework.
We first established a novel identity model based on identity entropy theoretically.
The individual conditional identity entropy and core identification set are defined to reveal the intrinsic mechanism of multivariate attribute collaborative identification.
Based on the proposed identity model, we propose a trustworthy identity tracing framework (TITF) with multi-attribute synergistic identification to determine the identity of unknown objects, which can optimize the core identification set and provide an interpretable identity tracing process.
Actually, the essence of identity tracing is revealed to be the process of the identity entropy value converges to zero.
To cope with the lack of test data, we construct a dataset of 1000 objects to simulate real-world scenarios, where 20 identity attributes are labeled to trace unknown object identities.
The experiment results conducted on the mentioned dataset show the proposed TITF algorithm can achieve satisfactory identification performance.
Furthermore, the proposed TIT task explores the interpretable quantitative mathematical relationship between attributes and identity, which can help expand the identity representation from a single-attribute feature domain to a multi-attribute collaboration domain. 
It indeed provides a novel perspective for realizing credible and interpretable identification techniques in real-world scenarios.
\end{abstract}




\maketitle

\section{Introduction}
During the investigation process of various criminal cases, the public security department uses identity tracing technology to confirm the identity of suspects. The specific task of identity tracing technology is to identify, from a wide range of data, the attributes needed to obtain information about a specific person, filter out important clues from massive data, and use attribute selection strategies to narrow down the investigation scope in order to obtain the identity tracing path of the target person. Currently, this work mainly faces the following problems: (1) Existing identity tracing models are still in the theoretical exploration stage and lack theoretical guidance\cite{shaheed2021systematic}; (2) In real scenarios, obtaining physiological information related to the case is complex, and the quality of collected data is low. Therefore, more than traditional single biometric attribute recognition technology is needed to meet the practical application requirements of law enforcement agencies. Multi-attribute collaborative identification\cite{mordini2012second} is currently the mainstream method, but there is a lack of research on how to design a multi-attribute collaborative identification model to improve the accuracy of identity recognition; (3) Filtering important clues and determining the investigation direction heavily rely on the professional ability of intelligence officers, and their experience level is an essential factor affecting the quality of intelligence analysis work. In a high volume of criminal and public order cases, a shortage of high-level criminal investigation experts leads to low investigation efficiency.
\begin{figure}
    \centering 
    \includegraphics[width=1.0\textwidth]{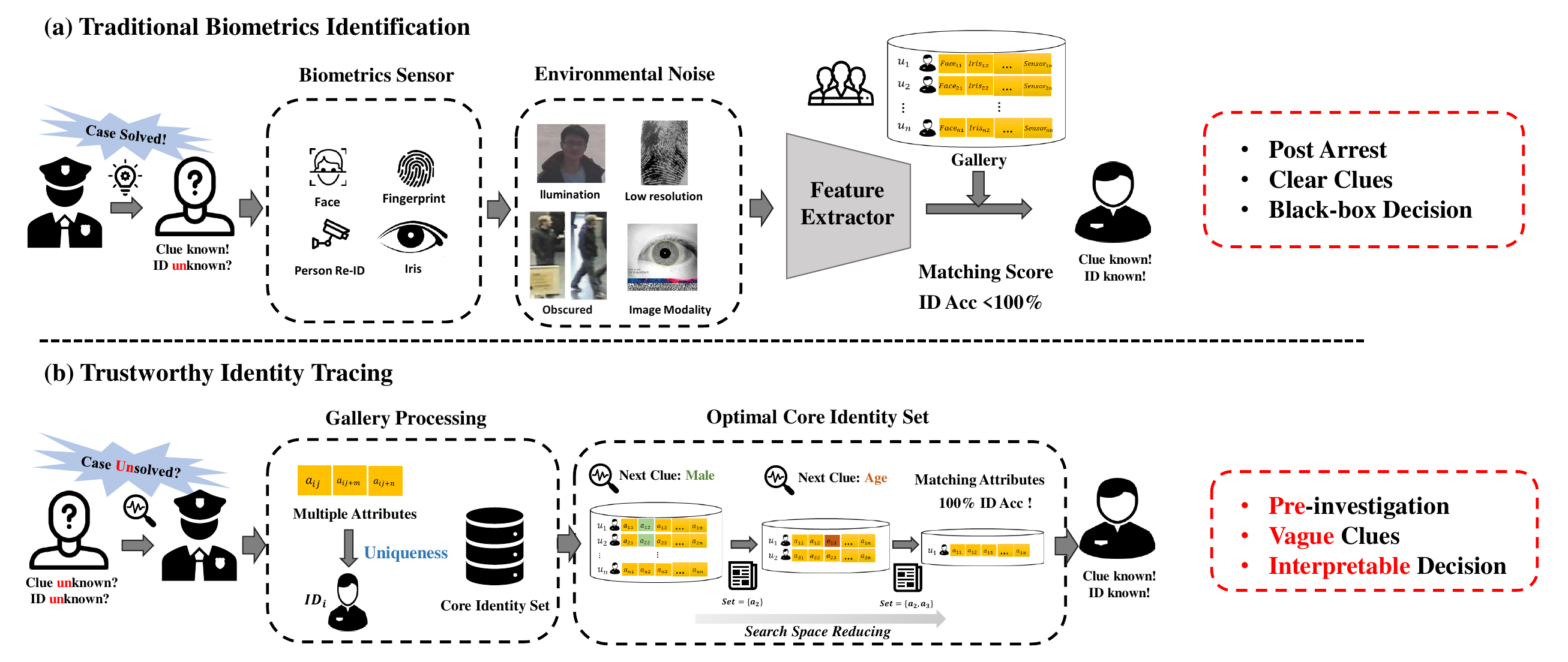}
    \caption{Illustrations of the proposed Trustworthy Identity Tracing task compared with traditional biometrics identification task.}
    \label{Figure1}
\end{figure}

Identity recognition works can be divided into two categories: single-attribute and multi-attribute collaborative recognition. 
Most single-attribute recognition based on images has developed from low-level image features to high-level semantic feature extraction based on deep learning\cite{liu2019high} and has continuously increased the amount of channel information to improve feature resolution and recognition accuracy\cite{kim2022adaface}. However, even the latest methods cannot achieve 100\%.
Researchers have focused on multi-attribute collaborative recognition to overcome the problems of low recognition accuracy and small application scope in single-attribute recognition\cite{walia2019robust}.
Existing studies have focused on fusion but have yet to consider whether there is an optimal feature combination for selection.
Current research needs to learn how to select the optimal feature combination for recognition, and researchers still need a deeper understanding of the underlying mechanisms of attribute collaboration.

Considering the introduction of multiple big data for identity recognition, in addition to the currently used biometric features, behavioral features, and other biological features, individuals' behavioral features and social features have specific recognition capabilitie\cite{srivastava2020words}, such as age, occupation, income, etc. Although the recognition ability is limited, it is easy to obtain. Multi-attribute collaborative identification can make up for the shortcomings of single-attribute recognition. The difference between attributes lies in whether they are possessed by the group, such as occupation, or only by individuals, such as DNA\cite{cao2010survey}. The contribution of attributes to identity recognition varies, so consider introducing a model to quantify the quantitative mathematical relationship between attributes and attributes, as well as between attributes and identity, to explore the essence of multi-attribute collaborative recognition and achieve optimal collaboration between attributes/attribute sets.

We summarize the contributions of this paper as follows:

\begin{enumerate}
	\item To our knowledge, we are the first to raise the task of Trustworthy Identity Tracing (TIT). Different from the existing identity recognition task (e.g. face recognition, Person Re-identification, fingerprint identification, etc.), our task minimize the search space most efficiently and support real-time updating of attributes to comprehensively give the optimal tracing direction. Unlike the followers studied before, trustworthy identity tracing can significantly reduce the efficiency due to the experience difference of investigators in practical applications.
	
	\item We verify the assumption that the relationship between the designed identity entropy and the attribute discriminability is significant. Furthermore, we prove the process of identity recognition is indeed equivalent to the process where the identity entropy value converges to zero.
	
	\item We first propose the efficiency problem of the identity tracing task, which considers the acquisition cost of attributes and mimics real-world scenarios. To solve the problem, we utilize the explored inherent relationship to construct an optimal core identifier set. Moreover, We devise a Trustworthy Identity Tracing Framework (TITF) to recognize object identity from collected multivariate data directly. And we evaluate its performance on a simulation dataset. Experiments of TITF) result in an efficient improvement compared to the traditional attribute mining-based method.
    \item We construct an Identity Tracing Dataset (ITD) containing a large-scale simulated dataset. The simulated dataset contains multivariate data from 1,000 objects. We believe the self-constructed datasets will benefit this field. We will release the datasets soon.
\end{enumerate} 

\section{Related work}
\subsection{Identity Identification}
The characteristics of identification mainly include biological characteristics and behavioral characteristics. Common attributes used for identity identification include face, palmprint, iris, gait, etc\cite{minaee2023biometrics}. In 2022, Fadi Boutros et al. proposed a face recognition algorithm based on elastic matching, which performed better than other methods in seven of the nine mainstream benchmark tests\cite{boutros2022elasticface}. Huikai Shao et al. proposed a weight-based metameric learning (W2ML) method for accurate open non-contact palm print recognition, improving accuracy by 9.11\%\cite{shao2022towards}.
In order to overcome the problems of low accuracy and small scope of application of single attribute identification, researchers focus on multiple identity identification\cite{walia2019robust,li2021practical}. The core of collaborative identification is how to effectively integrate multiple features, which can be divided into four basic levels\cite{bala2022multimodal}. Image layer (sensor) fusion: image data collected by various sensors are fused to form a new set of data sent to the subsequent feature extraction module. In 2012, Salim Chitroub proposed a recognition system based on FKP and FP, which fused at the image level and significantly improved the recognition accuracy compared with a single mode\cite{meraoumia2012multimodal,ryu2021continuous}. 
Furthermore, fusion is limited by the compatibility of the image. Feature layer fusion: Purohit Himanshu et al. proposed a feature-level multi-mode biometric feature fusion method in 2020, which improved the accuracy by 4.1\% compared with similar methods\cite{purohit2021optimal}. 
To sum up, the research focus of multi-modal biometric feature fusion has gradually shifted from single-level fusion to multi-level fusion\cite{cui2023multi,zhou2022multi,li2022multi} to improve the system's robustness and accuracy. 
In multi-attribute collaborative identification, the existing researchers have specified several features for fusion to improve the accuracy through model optimization\cite{zhou2019person,gao2021mso,xu2021multi}.
The research focuses on fusion but needs to consider whether there is an optimal combination of selected features. The existing researches need to learn how to select the optimal feature combination for identification, and researchers still need in-depth exploration of the internal mechanism of attribute synergy. At the same time, it is not easy to obtain some biometric features in practical applications, and the acquisition quality and application scenarios limit the effect. 

\subsection{Attribute Relationship in Identification}
In identity recognition research, some researchers focus on the physiological features of the subject, while others pay attention to other appearance features, such as clothing, accessories, and colors, as important bases for identity recognition.
In 2016, Matsukawa et al.\cite{matsukawa2016person} conducted fine-tuning on the pedestrian attribute dataset, combining attributes to distinguish similar groups of people. 
By combining multiple attributes, objects can be classified into different categories, thereby improving the performance of CNN features in pedestrian recognition and enhancing their discriminative ability.
In 2019, Lin et al.\cite{lin2019improving}proposed attribute based-pedestrian recognition (APR) network for re-identification using identity labels and attribute annotations. 
In 2020, Shi et al.\cite{shi2020person}proposed an attribute mining and reasoning (AMR) network for pedestrian retrieval tasks to discovered potential relationships, and generated more comprehensive pedestrian descriptions.
In 2021, Li Chao et al.\cite{li2021pedestrian}proposed a graph convolutional network for modeling semantic reasoning between attributes, demonstrating the relationship between combined simple attributes and overall characteristics for obtaining more robust and expressive features to improve pedestrian recognition. 
On the Market-1501 dataset\cite{zheng2015scalable}, the CMC-1 algorithm achieved a performance of 94.74\% and mAP of 87.02\%, while on the DukeMTMC-reID dataset\cite{zheng2017unlabeled}, CMC-1 achieved 87.03\% and mAP of 77.11\%, outperforming multiple state-of-the-art algorithms.

In summary, current researchers utilized attributes for identity recognition tasks, conducting multiple correlation analyses to expand the scale of identity attributes and achieve identity collaboration identification. After introducing multiple attributes, most existing models perform correlation analyses on all attributes, combine simple attributes to obtain stronger recognition ability and enhance model representational power. However, due to some semantic overlap between certain attributes, using prior knowledge to select multiple attributes and precisely calculate their representational power before and after combination, prioritizing attributes or attribute combinations with strong representation capabilities for the association, can maximize the role of attribute associations.

\section{Definiton and Motivation}{\label{definition}}

\subsection{Motivation}
The proposed novel Trustworthy Identity Tracing is inspired by the process of solving cases by police.
In real-world scenarios, the police always need to analyze various related clues and information of suspects to launch an investigation.
Before the case is solved, the identity of the suspect can't be confirmed and the traditional biometrics identification doesn't work.
We find that the police with rich experience can effectively filter important clues and shorten investigation time and vice versa.
Thus, we aim to construct an efficient trustworthy identity tracing system to automatically select more important attributes as more valuable clues, which can help the police narrow the investigation scope and shorten the case time.

\subsection{Problem Definition } \label{motivation}

In order to reveal the intrinsic mechanism of multi-attribute synergistic identification, we established an identity model based on identity entropy, and defined conditional identity entropy to quantify the process of individual identity recognition. Based on our proposed identity model, we define identity tracing to express the process of discovering the real identity of an object through the selection and collection of attributes of which to be tested.
The previously described application scenarios were transformed into the following mathematical model:

To sum up, for $U_{N} $, a collection of $N$ objects, $N$ objects are completely distinguished by $M$ different attributes. $M$ attributes $A_{i}=\left ( i=1,2,\cdots ,M \right ) $ , the $i^{th}  $ attribute $A_{i}$ has $k_{i}$ possible values, denoting $A_{i}\in \left \{ a_{i1},a_{i2},\cdots \cdots ,a_{ik_{i} } \right \}$ , probability distribution $P=\left \{ p\left ( a_{i1}  \right ), p\left ( a_{i2}  \right ) ,\cdots \cdots ,p\left ( a_{ik_{i} }  \right )\right \} $ , $p\left ( a_{ij}  \right )=\frac{\left | a_{ij}  \right | }{\left |U_{N}   \right | } \left ( 1\le i\le M,1\le j\le k_{i}  \right )  $. Where $\left | a_{ij}  \right | $ denotes the number of objects whose attribute value is $a_{ij}$ among this attribute category, $\left | U_{N}  \right | $ denotes the size of this search space, and the $N$ objects of the dataset $U_{N}$ and the $M$ attribute values corresponding to each object totaling $M\times N$ attribute values are known.

\begin{myDef}
 \label{definition 1}
    (Individual identity entropy).The uncertainty of an object $U_{x} $ is defined as the object's individual identity entropy $H\left ( U_{x}  \right )  $, and $n$ is the current search range of object $U_{x}\left ( n\le N \right ) $.
    \begin{equation}\label{(1)}
     H\left ( U_{x}  \right )=\log_{2}{n}  .
    \end{equation}
    Under the conditions that the attribute set $A$ is known, The identity entropy of an individual object is defined as the individual conditional identity entropy, denoted as $H\left ( U_{x}\mid A  \right ) $, $H_{0}\left ( U_{x}  \right )  $ is the initial identity entropy, and $I\left ( A \right ) $ is the discriminability of attribute set $A$.
    \begin{equation}\label{(2)}
    H\left ( U_{x}\mid A  \right ) = H_{0} \left (  U_{x} \right ) - I\left ( A \right ) .   
    \end{equation}
    Since in the process of discussing identity issues, what we call identity has a certain scope, so usually the individual identity entropy we calculate is the individual conditional identity entropy.   
\end{myDef}

\begin{myDef}
    \label{definition 2 }
	(Identity Tracing).For an unknown object $U_{x}$ in a known set $U_{N}$, $m$ attributes of the object $U_{x}$ are known, and the next proposed collection of attribute category $A_{x} $ is sought to maximize the probability that the search range decreases, and gradually optimize the path of identity tracing to finally determine the identity of the object $U_{x}$.
\end{myDef}

\begin{myDef}
    \label{definition 3 }
    (Core Identificaiton Set).The set of attributes that uniquely identifies an object and whose arbitrary true subset does not identify the object is called the core identificaiton set of that object.
    The sufficient and necessary conditions for a set of attributes $A$ to be the core identification set for object $U_{x}$ are as follows: (1) The attribute set A can uniquely identify the object $U_{x}$. (2) None of the proper subsets of attribute set A can uniquely identify any object within the search space.
\end{myDef}

\begin{myDef}
    \label{definintion 4 }
    (Attribute Discriminability).
 The magnitude of an attribute's impact on object identity discriminability is referred to the attribute discriminability. 
 
 Attributes $A_{i}=\left ( i=1,2,\cdots ,M \right ) $  has $k_{i}$ possible values, denoting $A_{i}\in \left \{ a_{i1},a_{i2},\cdots \cdots ,a_{ik_{i} } \right \}$ , probability distribution $P=\left \{ p\left ( a_{i1}  \right ), p\left ( a_{i2}  \right ) ,\cdots \cdots ,p\left ( a_{ik_{i} }  \right )\right \} $. The calculation formula for attribute discriminability is represented as:
\begin{equation}\label{(3)}
    I\left ( a_{ij}  \right ) =- \log_{2}{p\left ( a_{ij}  \right ) } 
\end{equation}
When attribute $a_{x} $ is known, the magnitude of another attribute $a_{y} $'s impact on object identity identification is referred to  as the conditional discriminability of attribute $a_{y} $. 
The calculation formula for attribute conditional discriminability is represented as :
\begin{equation}\label{(4)}
    I\left ( a_{y} \mid a_{x}  \right ) =-\log_{2}{p\left (  a_{y} \mid a_{x}\right ) } 
\end{equation}
\end{myDef}

\section{Methodology}
\begin{figure*}
   \centering
    \includegraphics[width=1.0\textwidth]{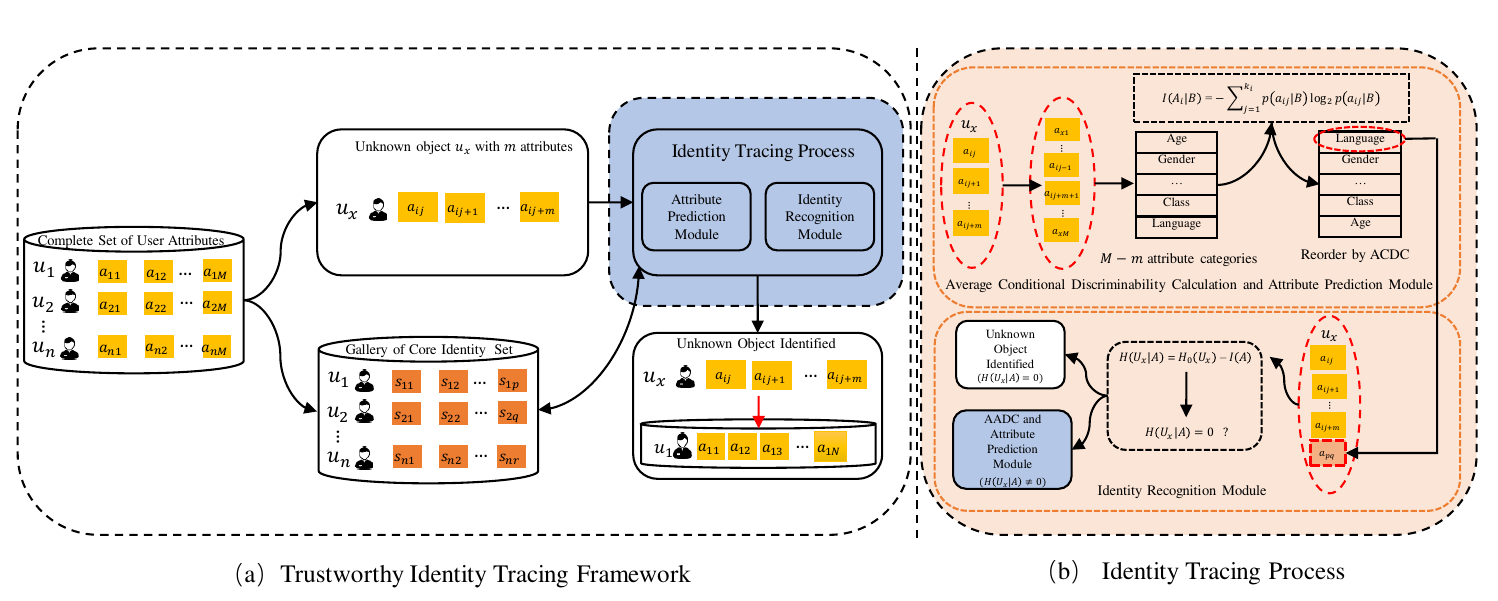}
    \caption{Illustrations of our proposed novel multi-attribute synergistic identification framework in the trustworthy identity tracing task.}
    \label{Figure2}
\end{figure*}

The scenario definition for identity traceability task by public security organs is as follows: During the investigation process of various criminal cases, based on various identity information and related information of an unknown object already obtained, the next optimal investigation direction is determined within a specific range by selecting and collecting the identity attributes of the target object, in order to narrow the investigation scope and ultimately discover the true identity of the object.

The problem definition is as follows: Given an unknown object $U_x$ in the known set $U_N$, and given $m$ attributes of the object $U_x$, find the next attribute category $A_x$ to collect that maximizes the probability of reducing the search scope the most, and gradually optimize the identity trace path, ultimately determining the identity of the object $U_x$

Then, we will introduce our proposed model trustworthy identity tracing framework (TITF), and the illustration of TITF is shown in \ref{Figure2}. Overall, It consists of three key components: (1)an core identification set calculation module to compute sets that can completely distinguish an object from other objects ; (2)an average conditional discriminability calculation and attribute prediction module to compute the optimal attribute that can minimize the scope of detection; (3) an identity recognition module to identify the object; 

\subsection{Core Identification Set Calculation Module}
\subsubsection{Core Identification Set}
Given an unknown object $U_x$ in the known set $U_N$, if there exists such an attribute or attribute set A that makes $H\left ( U_{x}\mid A  \right )=0$, then the attribute or attribute set A can distinguish the object $U_{x}$ from other individuals. We refer to such a set as the identification set for the object $U_{x}$.

If set $A$ is the identification set of object $U_{x}$, then for set $B$, if set $A$ is a subset of set $B$, set $B$ is also the identification set of object $U_{x}$. The above conclusion corresponds to the reality that no object has a unique identiofication set, and the composition of identification sets may not be the simplest, so in the task of identity recognition, we should pursue a more concise identificaiton set to achieve recognition, and to exclude the "redundant" attributes in the identification set. Thus, we give the definition of the core identification set.

The set of attributes that uniquely identifies an object and whose arbitrary true subset does not identify the object is called the core identificaiton set of that object

The sufficient and necessary conditions for a set of attributes $A$ to be the core identification set for object $U_{x}$ are as follows: (1) The attribute set A can uniquely identify the object $U_{x}$. (2) None of the proper subsets of attribute set A can uniquely identify any object within the search space.
\subsubsection{Optimal Core Identification Set}
Considering that some attributes with high identification contribution, such as DNA (disregarding very special cases), ID number (real), etc. can individually identify the object, it is concluded that the core identification set of an object may not be unique in context, i.e., the composition of the core identification set may be more than one under the same search space.   

Due to the diversity of the composition of the core identification set and the difference in the disciminability of attributes, we propose the concept of optimal core identification set.

To construct an optimal core identification set for an object,  the highest average conditional discriminability of the attribute needs to be selected for each attribute selection. The algorithm flow is shown in Algorithm\ref{algorithm 1 }.

\begin{algorithm}[!h]

  \SetAlgoLined
  \KwIn{all attributes date of all objects: $M$ attributes of $N$ objects, partial known attributes of target: $m$ attributes, object $U_{x} $ }
  \KwOut{optimal core identification set :set $A$ }
  
  Search space: $n = N$, $A =\emptyset$, $m=0$, initial identity entropy $H_{0}\left ( U_{x}  \right)=\log_{2}{N}  $, according to formula $H\left ( U_{x}\mid A  \right ) = H_{0} \left (  U_{x} \right ) - I\left ( A \right ) $, calculate the individual conditional identity entropy $H$ \;
  
  \If {H \textgreater 0}
  { 
     \For{$H \textgreater 0,m\le M$}
     {
        Based on the formula $\ref{(1)}-\ref{(4)}$ to calculate the average conditional discriminability of the remaining $M-m$ attributes, the attribute with the highest average conditional discriminability is added to set $A$\;
        Then calculate the individual conditional identity entropy $H$ of object $U_{x} $, $m=m+1$ \;        
     }
  }
  The set $A$ is the optimal core identification set\;
  \caption{Calculating the Optimal Core Identification Set}\label{algorithm 1}
\end{algorithm}

\subsection{Average Conditional Discriminability Calculation and Attribute Prediction Module}
We have discussed the concepts of identity entropy and conditional identity entropy, and defined the set that can uniquely identify an object without containing redundant attributes as the core identification set for the object. However, regarding the contribution magnitude of attributes or attribute sets in identity identification, only qualitative analysis has been conducted. The following section presents a quantitative analysis of attribute discriminability.
\subsubsection{Relationship of Attribute Discriminability and Core Identificaiton Set}

For the object $U_{x} $ with a current search space size of $n$, its initial identity entropy is $H_{0}\left ( U_{x}  \right ) =\log_{2}{n} $. When introducing an attribute $a_{x} $ for this object, there are $m$ objects in the current search space with the same attribute value as $a_{x} $, resulting in the search space reducing to $m$ objects, with all objects having different attribute values being eliminated. Therefore, after introducing attribute $a_{x} $, the conditional identity entropy of object $U_{x} $ becomes $H\left ( U_{x} \mid a_{x}  \right )=\log_{2}{m}   $. According to $\left ( \ref{(2)} \right ) $, the discriminability contribution of attribute $a_{x} $ is calculated as $I\left ( a_{x}  \right ) =H_{0}\left ( U_{x}  \right )- H\left (U_{x} \mid a_{x}    \right )=\log_{2}{\frac{n}{m} } $.Returning to the original search space of size $n$, where there are $m$ objects with attribute value $a_{x} $, according to $\left ( \ref{(3)} \right ) $, the discriminability of attribute $a_{x} $ is given by $I\left ( a_{x}  \right ) =- \log_{2}{p\left ( a_{x}  \right ) } =\log_{2}{\frac{n}{m} } $.In conclusion, under this model, the discriminability magnitude of an attribute and its contribution to identity identification are numerically equal.

\subsubsection{ Average Conditional Discriminability of Attribute Category}
The average contribution of an attribute category $A_{i} $ to identity recognition under the condition of known attribute set $B$.
The calculation formula for attribute conditional discriminability is represented as:
\begin{equation}\label{(5)}
    I\left ( A_{i}\mid B  \right ) =-  {\textstyle \sum_{j=1}^{k_{i} }} p\left ( a_{ij}\mid B  \right )\log_{2}{p\left ( a_{ij}\mid B  \right ) }
\end{equation}
The attribute set $A_{k} $ represents a set containing $k$ attributes, where $a_{i}\left ( i=1,2,3,\cdots k \right )$   denotes the $i^{th} $ attribute in the set. The calculation formula for the discriminability of attribute set $A_{k} $ is as follows:  
\begin{equation}\label{(6)}
I\left ( A_{k}  \right ) =
\begin{cases}
- \log_{2}{p\left ( a_{1}  \right ) } ,\quad &k=1 \\
I\left ( A_{k-1}  \right ) +  I\left ( a_{k} \mid A_{k-1}  \right ) \quad &1< k\le m   
\end{cases}
\end{equation}

\begin{proof}\renewcommand{\qedsymbol}{}
    The set $A_{k}$ contains $k$ attributes denoted as $A_{k}$.When $k = 1$, according to $\left ( \ref{(3)}\right ) $, we can infer that $I\left ( A_{1}  \right )=I\left ( a_{1}  \right )=- \log_{2}{a_{1} } $.When $k = 1$, the equation is proved.For $1< k\le m$, based on $\left (\ref{(4)}\right ) $ and $\left ( \ref{(5)} \right ) $, we can deduce that $I\left ( A_{2}  \right ) =I\left ( a_{1}a_{2}   \right ) =I\left ( a_{2}\mid A_{1}   \right ) +I\left ( A_{1}  \right ) $,Now, assuming $k = c \left ( 1< c\le m-1 \right ) $ when the equation holds true,we have $I\left ( B_{c}  \right ) =I\left ( b_{1}b_{2}\dots b_{c}  \right ) =I\left ( b_{c}\mid B_{c-1}  \right )+I\left ( B_{c-1}  \right ) $.When $k = c + 1$, according to $\left ( \ref{(5)} \right )$  , we have$I\left ( B_{c+1}  \right )  =I\left ( b_{c}\mid B_{c-1}  \right )+I\left ( B_{c-1}  \right )$.
    In conclusion, for $1\le k\le m$, the equations hold true.
\end{proof}

Through analysis and reasoning, we have clarified the equational mathematical relationship between identity uncertainty and attribute discriminability. Building upon this foundation, an identity model based on identity entropy is established, unveiling the fundamental mechanisms of identity representation and clarifying the mathematical essence of identity. Complete quantitative calculation methods are derived, pointing out the essence of multi-attribute collaborative identification and providing theoretical support for identity tracing.

\subsection{Identity Recognition Module}
This algorithm uses the idea of constructing the optimal core identifier set to explore the optimal trace path.

When tracing the identity of an unknown object, it is not possible to calculate the discriminability or conditional discriminability of specific attributes. Therefore, only the conditional average discriminability of attribute categories can be used as the basis for decision-making.

Since the average discriminability is used to evaluate the contribution of an attribute to identity discrimination, for a single tracking task, the attribute may not be the optimal solution of the attribute to be obtained in this step. However, after a sufficient number of tracking tasks are performed, the average discriminability of a specific attribute value will be closer and closer to the average discriminability of the attribute category, so the optimal attribute category recommended by this method is the probabilistic optimal attribute category to be obtained.
The algorithm flow is shown in Algorithm \ref{algorithm 2  }.
\begin{algorithm}[!h]
  \SetAlgoLined
  \KwIn{all attributes date of all objects: $M$ attributes of $N$ objects, partial known attributes of target: $m$ attributes}
  \KwOut{target identity:ID,trace path:set $A$ }
    Search space: $n = N$, $A = \left\{m\quad attributes\right\} $, according to formula $\left ( \ref{(5)}\right ) $,calculate the average conditional recognition of $M-m$ attributes of $N$ objects under any conditions\;
  Set $A$ is matched with $M$ attributes to get the current search space $n$\;
  \If {n \textgreater 1}
  { 
     \For{n \textgreater 1}
     {
        According to the average conditional discriminability of attributes, the attribute with the highest average discriminability is obtained, and the attribute is added to set $A$\;
        Set $A$ is matched with $M$ attributes to get the current search space $n$\;
     }
  }
  Output target identity: ID, set $A$. The attributes in set $A$ is the trace path\;
  \caption{TITF}\label{algorithm 2 }
\end{algorithm}

\section{Experiment}
In this experiment, we conducted experiments on public data sets to answer the following questions:
\begin{itemize} 
	\item \textbf{RQ1} Is the object core identification set unique in the observation space?
	\item \textbf{RQ2} In the same observation space, is there any difference in the discriminability of the same attribute for different objects?
	\item \textbf{RQ3} Is there any difference in the discriminability of the same attribute for the same object in different observation spaces?
	\item \textbf{RQ4} In an ideal state, how efficient is the trustworthy identity tracing framework(TITF)?	
\end{itemize}

\subsection{Simulation Dataset}\label{dataset}
The dataset used in this study originates from the USCensus1990raw data set\cite{uci1990us}. The data is derived from the results of the 1990 United Census in the United States, and the identity attributes information in the data set was obtained through questionnaires. The dataset contains a total of 2,458,285 individuals, each with 68 attributes. Due to the presence of attributes with low discriminatory power or no discriminatory power (where most individuals have the same attribute values) in the USCensus1990 data set, a selection of 20 different categories of attributes was made based on consideration of differentiation of attributes and generalizability in practical applications. All personal information was anonymized to protect privacy.In the whole paper, we only reported the overall statistical data of this dataset, without revealing any identifiable personal information. We can ensure that our data is processed in a manner that complies with the legal requirements of the country in which the data subject belongs.

\subsection{Comparison Results}
In practice, when the identity tracing task is carried out without theoretical guidance and experience, there is no clear direction for the next acquisition of attributes, so the random search method is selected to simulate the practical application method and serve as the comparison method of this method.

\textbf{Baseline:} Random search method\cite{zabinsky2009random}: in the process of identity tracing, the attribute category is randomly selected as the next attribute to be obtained. Because the target object can be distinguished in the data set, the target object can always be identified after obtaining enough attributes.

\subsection{Experiment Settings}
In this study, the search time $T$ is used to measure the performance of the method.In the process of identity traceability, the time to continuously obtain attributes and finally confirm the target object is search time T, and the unit is milliseconds, which is used to evaluate the efficiency of the method.

\subsection{Assumption Verification}
Firstly, we verify the diversity of the core identification set in the proposed identity entropy model entropy(Q1), and put forward two important assumptions: there are differences in the discriminability of the same attribute of different objects in the same observation space(Q2); There are differences in attribute discriminability of the same object in different observation spaces(Q3).The verification of them provides new knowledge in this field and supports us to design the framework TITF.

For Q1. We select 5000 individuals and 20 attribute categories from the public data set US Census Data as the experimental data, and divide the 5000 individuals in the data set into 10 groups with 500 individuals in each group. Randomly select 10 individuals and put them into 10 groups, calculate the core identification set of the object in different groups, and record the number of the core identification set of the target object in each group respectively.

\begin{figure}[pos=htp]
    \centering
    \includegraphics[width=0.6\textwidth]{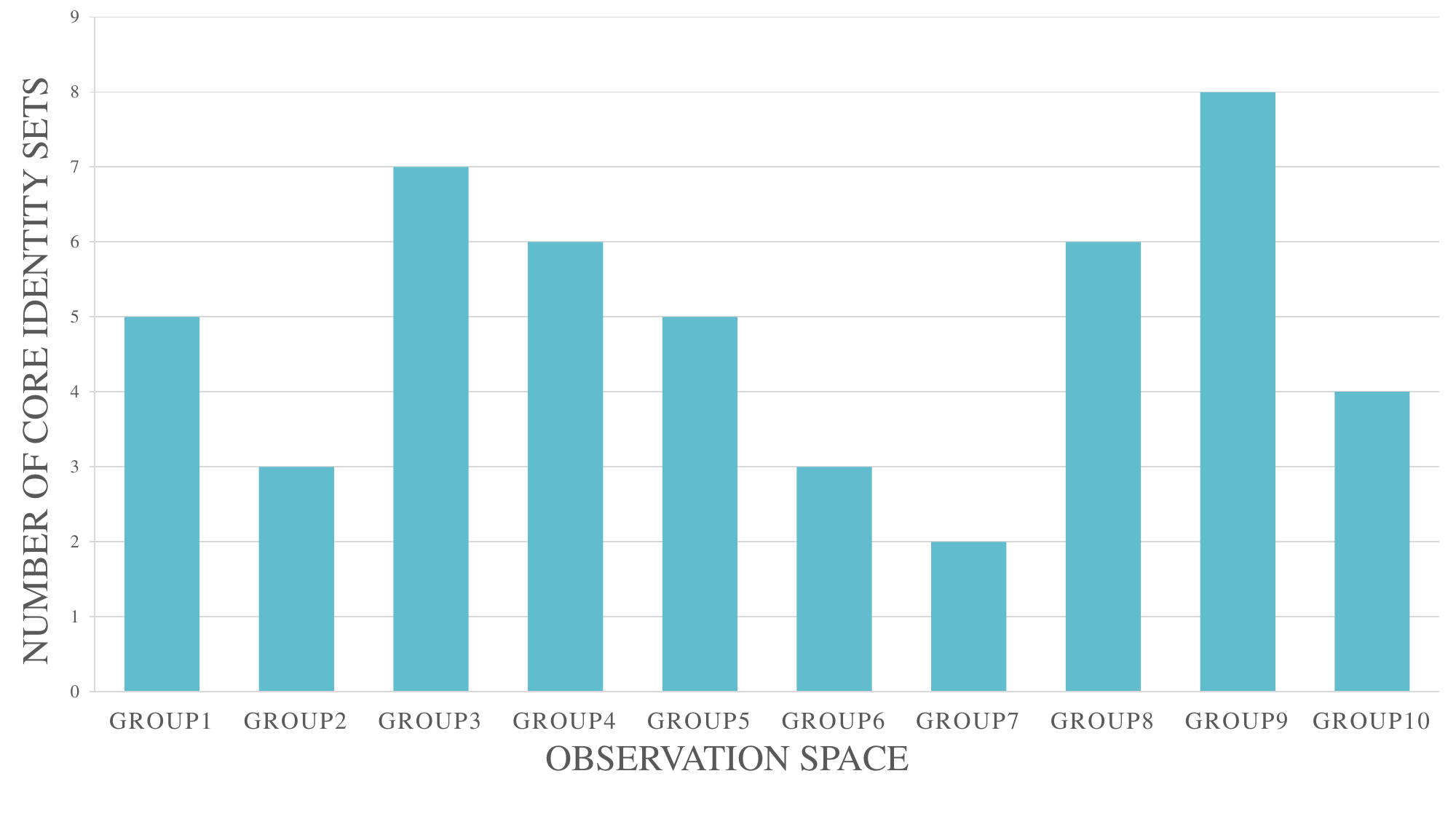}
    \caption{Number of core identification sets in different observation space for different individuals.}
    \label{figure 3 }
\end{figure}

As can be seen from Figure \ref{figure 3 }, for different objects in different observation spaces, any object has one or more core attribute sets corresponding to it. For each object, when adding different attributes to solve the core identification set, the composition of the final core identification set may be different, which makes the diversity of the core identification set possible.

For Q2, we select a data set containing 1000 individuals and 20 attribute categories from the public data set US Census Data, and divide the 1000 individuals into 10 groups, each with 100 individuals in an observation space. We input the grouped data set into the identity entropy model, calculate the attribute discriminability of 20 attribute categories of all 1000 individuals, and compare their differences.

\begin{figure}[pos=htp]
    \centering
    \includegraphics[width=1.0\textwidth]{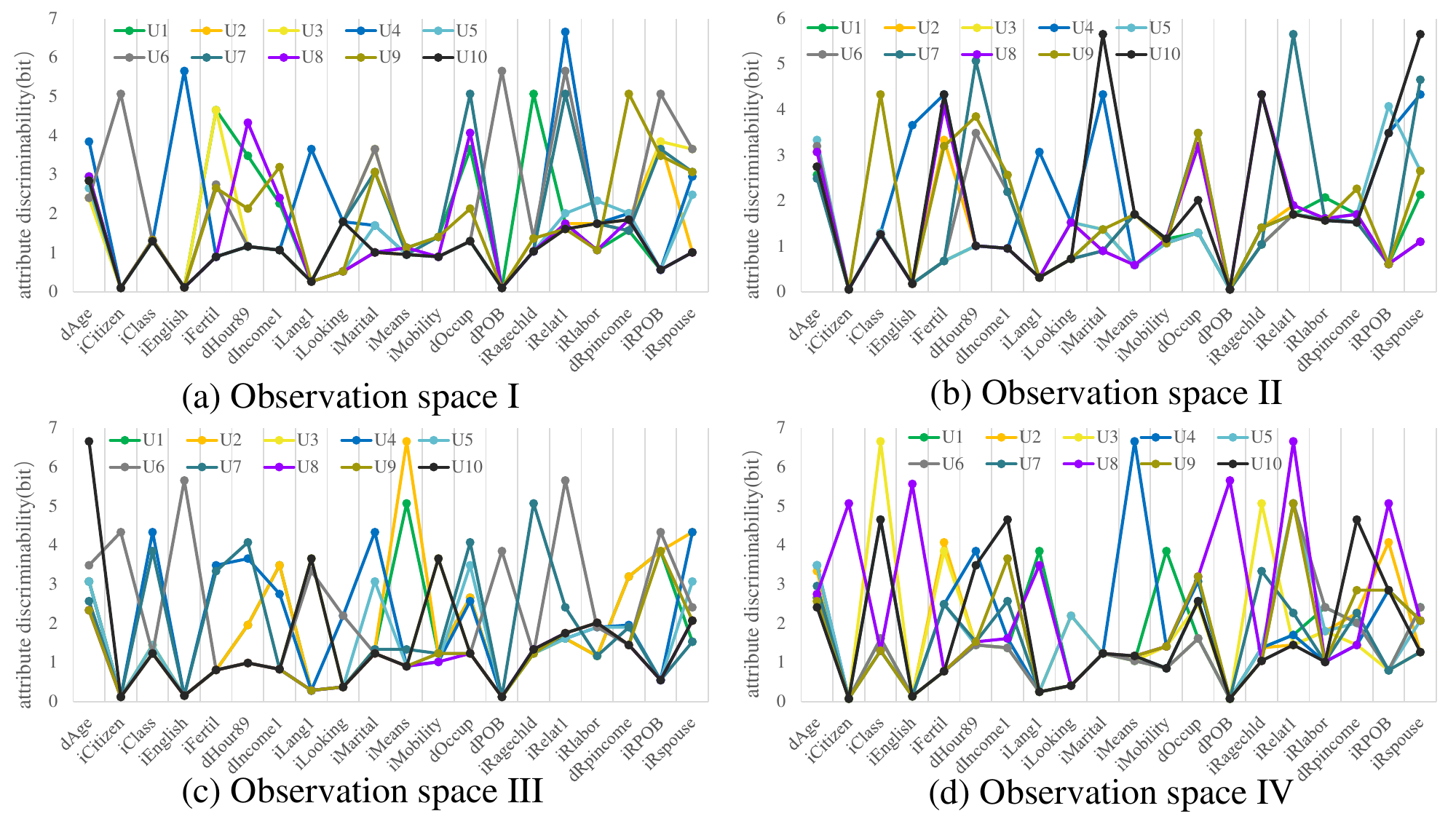}
    \caption{Attribute discriminability of different objects in different observation Spaces.}
    \label{figure 4 }
\end{figure}

As can be seen from Figure \ref{figure 4 }, for different objects in the same observation space, there are obvious differences in the discriminability of attributes of the same attribute category. According to the attribute discriminability formula, if the attribute value distribution of the same attribute category is different, the contribution to different object recognition is also different. In particular, if the attribute class has the same value for all objects in the observation space, the attribute class is not discriminating.

\begin{figure}[pos=htp]
    \centering
    \includegraphics[width=0.6\textwidth]{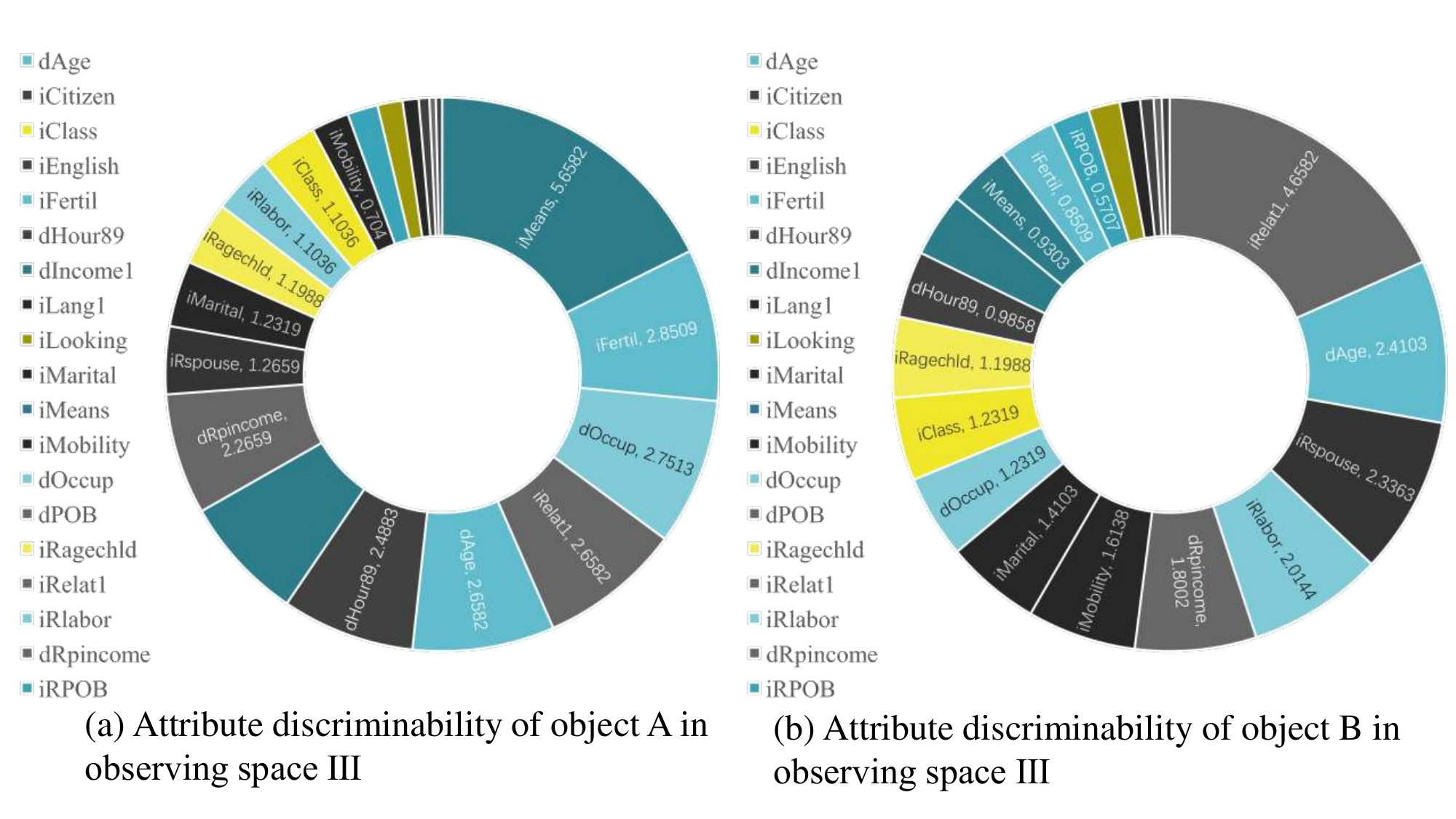}
    \caption{Attribute discriminability of different objects in the same observation space.}
    \label{figure 5 }
\end{figure}

Select two objects from the same observation space, and arrange all their attributes according to the attribute discriminability, as shown in Figure \ref{figure 5 }. In the process of expressing the identity of different objects, choosing different attribute categories can improve the conciseness of identity expression. There is a personalized optimal solution for the identity expression of different objects instead of a uniform template.

\begin{figure}[pos=htp]
    \centering
    \includegraphics[width=1.0\textwidth]{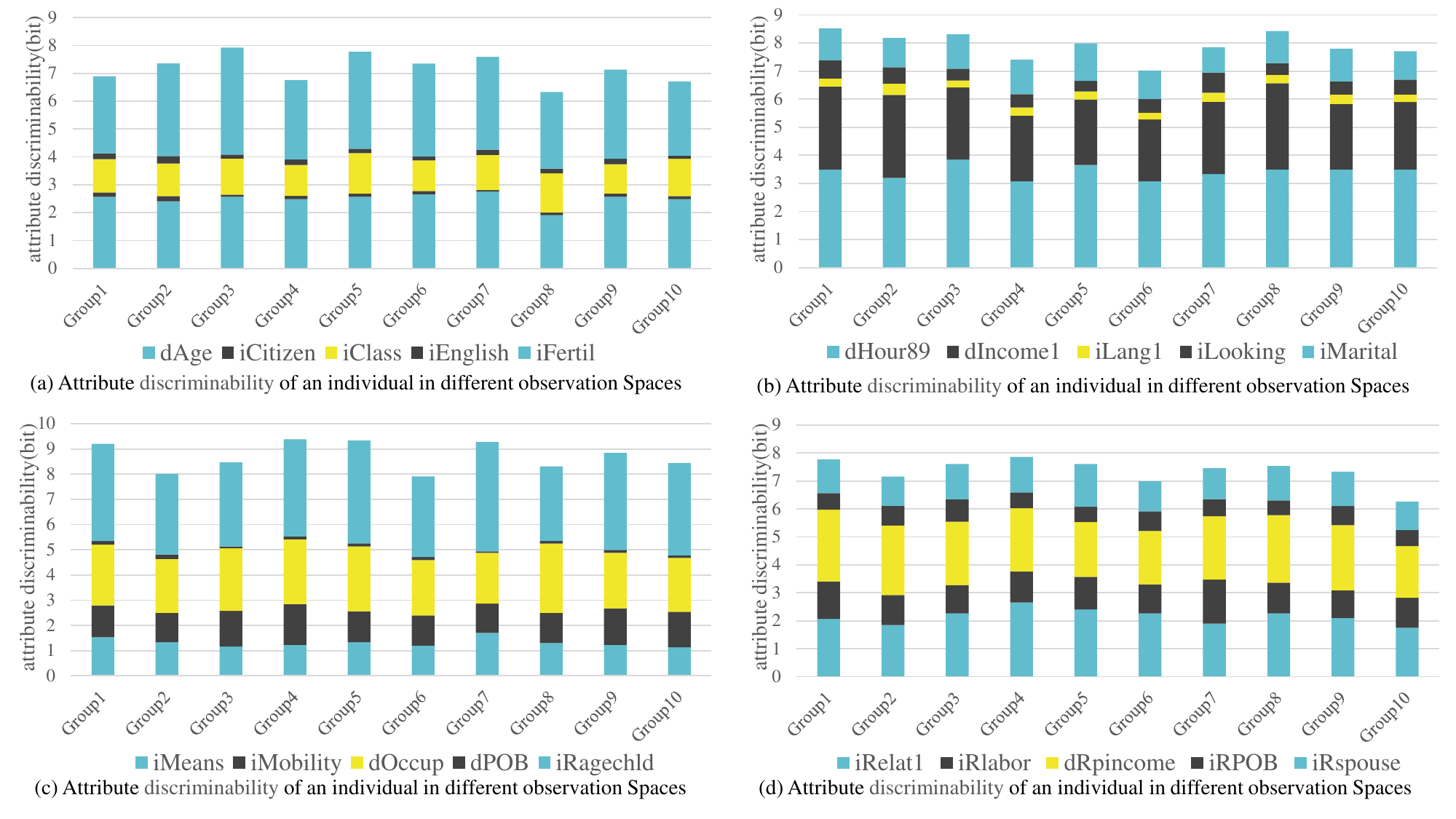}
    \caption{Different attribute discriminability of the same object in different observation space.}
    \label{figure 6 }
\end{figure}

For Q3, we select 1,000 individuals with 20 attribute categories from the public data set US Census Data, and divide the 1,000 individuals into 10 groups on average. We select one individual from the data set as the target object and put it into 10 groups respectively. We input the data set and the target object into the identity entropy model, calculate the discriminability of 20 attribute categories of the same target object in different observation spaces, and compare their differences.

As can be seen from Figure \ref{figure 6 }, In different observation spaces, although the attribute value of the same attribute category for the same object is the same, the probability distribution of the attribute value is different due to the different observation spaces. According to the calculation formula of attribute discriminability, the contribution of the same attribute to the identity discrimination of the same object in different observation spaces is also different. In different observation spaces, there is also a personalized optimal solution for the identity expression of the same object, rather than a fixed set of attributes. Combined with Q2, we have the following conclusions: We should seek an adaptive personalized method for identity tracing tasks of different scenes and different objects.

\subsection{Performance of TITF}
For Q4, We select 5000 individuals and 20 attribute categories from the public data set US Census Data. Input the data set as a training set into the identity traceability system, and preprocess the data. In the case of attribute missing $S$=20\%, 40\%, 60\% and 80\% (artificially deleting some data to simulate partial known attributes in practical application), partial known attributes of the object are used as input, and the unknown objects are traced by Ours and Baseline respectively. 
\textbf{Baseline: }for the random search method\cite{zabinsky2009random}, each unknown object is traced back to its identity for 20 times, and the average time consumption is taken as the time consumption of this unknown object, and the average time consumption of 100 objects is taken as the time consumption of the random search method. 
\textbf{Ours:} For TITF, each object is traced only once, and the time-consuming average of 100 objects is taken as the time-consuming of this method.

\begin{figure}[pos=htp]
    \centering
    \includegraphics[width=0.6\textwidth]{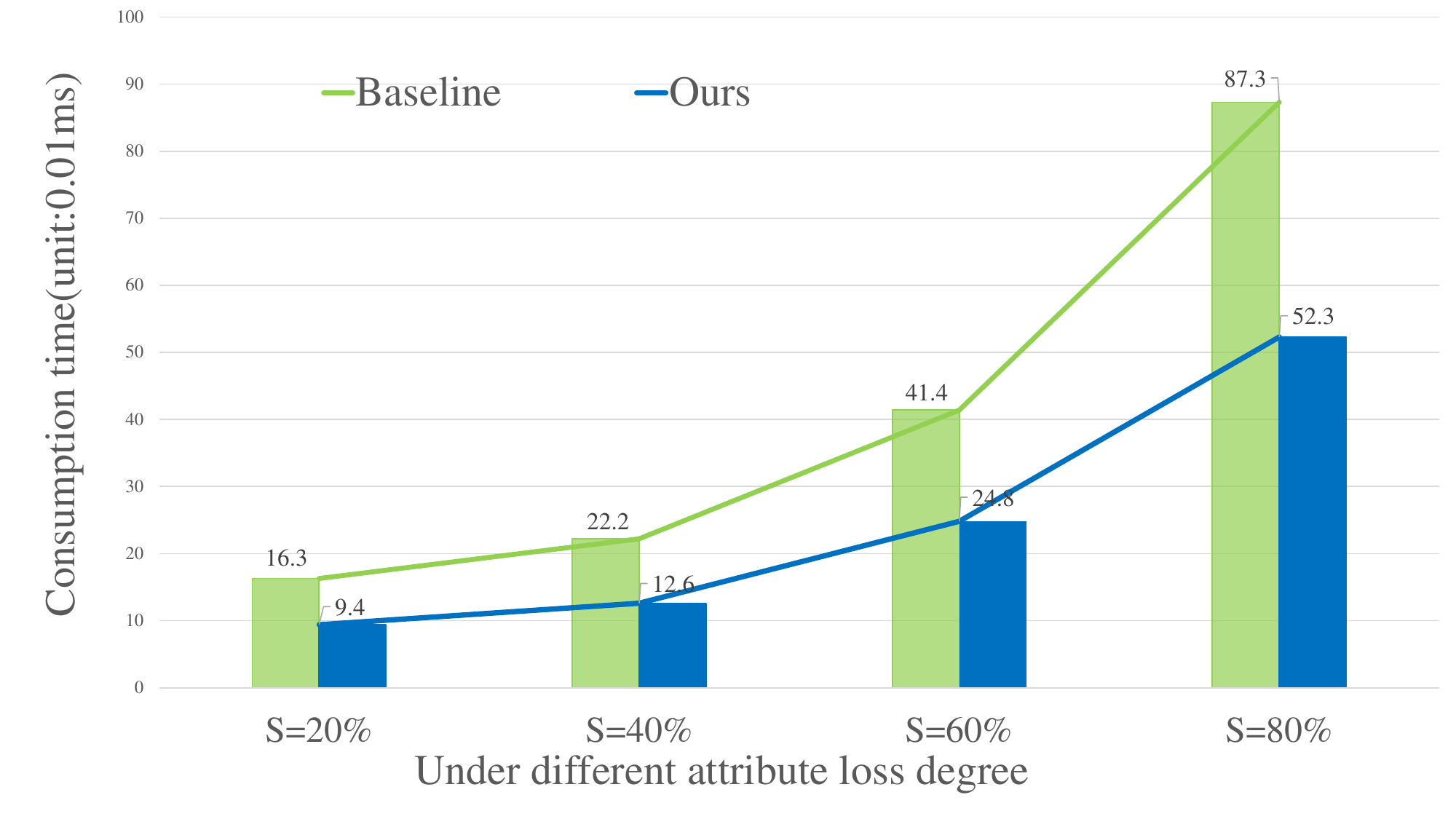}
    \caption{Average consumption time for trace identity.}
    \label{figure 7 }
\end{figure}

Figure \ref{figure 7 } show the average time-consuming of the baseline and ours method when the attribute is missing $S$=20\%, 40\%, 60\% and 80\%, regardless of the difficulty of attribute acquisition. Regardless of the difficulty of attribute acquisition, compared with the baseline, our proposed method reduces the time consumption of identity traceability by 41.44\% on average.
For the baseline, it is easy to obtain attributes with low conditional discriminability and even redundant attributes, which leads to a longer time-consuming identity traceability. For TITF, according to the difference of the average conditional discriminability of attributes, the attribute category that is most likely to reduce the search scope most can be selected in each selection game. By introducing the average conditional discriminability, the acquisition of redundant attributes can be avoided to the maximum extent, and the performance is significantly improved compared with the traditional method without theoretical guidance or lack of experience.

\section{Conclusion and Future Work}
In this paper, we first propose the novel Trustworthy Identity Tracing (TIT) task.
The novel identity model is established theoretically to real the instinct mechanism of multivariate attribute collaborative identification in the process of identity tracing.
The multi-attribute synergistic identification-based Trustworthy Identity Tracing Framework (TITF) is proposed to prove an interpretable identity tracing process.
We also constructed a new simulation dataset of 1000 objects to evaluate the effectiveness of the proposed method.
In the future, we will evaluate the efficiency performance of the identity tracing process on more complex scenarios with noise or adversarial labeled data, which can mimic the needs of the real world.   



\bibliographystyle{cas-model2-names}

\bibliography{cas-refs}


\end{document}